\documentclass[pdftex,preprint,showpacs]{revtex4}
\usepackage[pdftex]{graphicx}
\usepackage{dcolumn}
\usepackage{bm}

\def\Journal#1#2#3#4{{#3} {\sl #1} {\bf #2} {#4}}

\def\NP{{ Nucl. Phys.} }
\def\NPA{{ Nucl. Phys. A}}
\def\PLB{{ Phys. Lett.}  B}
\def\PL{{ Phys. Lett.}}

\def\PRP{{ Phys. Rep.}}
\def\PRL{ Phys. Rev. Lett.}
\def\PR{{ Phys. Rev.}}
\def\PRD{{ Phys. Rev.} D}

\def\ZPC{{Z. Phys.} C}
\def\EPJ{{Eur. Phys. J.}}
\def\EPJC{{Eur. Phys. J.} C}

\def\CPC{Comput. Phys. Commun.}
\def\JHEP{J. High Energy Phys.}

\def\ra{\rightarrow}

\def\be{\begin{equation}}
\def\ee{\end{equation}}
\def\bea{\begin{eqnarray}}
\def\eea{\end{eqnarray}}

\def\qbar{{\bar q}}
\def\ubar{{\bar u}}
\def\dbar{{\bar d}}
\def\sbar{{\bar s}}

\def\NP{{ Nucl. Phys.}}
\def\ANP{{Adv. Nucl. Phys.}}
\def\CPC{{Comput. Phys. Commun.}}

\begin{document}




\title{Strange sea distributions of the nucleon}
\author{H. Chen}
\affiliation{School of Physical Science and Technology, Southwest University, Chongqing 400715, People's Republic of China}
\author{F.-G. Cao and A. I. Signal}
\affiliation{Institute of Fundamental Sciences PN461, Massey University,
Private Bag 11 222,  Palmerston North, New Zealand}

\begin{abstract}
The strange and antistrange quark distributions of the nucleon are less constrained by experimental data than the non-strange quark sea. 
The combination of light quark sea distributions, $\Delta (x)=\dbar(x)+\ubar(x)-s(x)-\sbar(x)$, 
originates mainly from non-perturbative processes and can be calculated using non-perturbative models of the nucleon.
We have calculated $\Delta(x)$ using the meson cloud model, which, when combined with the relatively well known 
non-strange light antiquark distributions obtained from global analysis of available experimental data, enables us to make new estimates of the total strange sea distributions of the nucleon and the strange sea suppression factor.

\end{abstract}

\pacs{14.20.Dh, 12.39.Ba}
\maketitle


\section{Introduction}

The strange and antistrange quark distributions (hereafter referred to as strange sea distributions)
of the nucleon are of great interest since these distributions are important
for many processes in high-energy hadron collisions. 
For example, a precise understanding of the cross-section for $W$ production at the Large Hadron Collider (LHC)
depends on the strange sea distributions in the small $x$ region.
However, the strange sea distributions are not well determined compared with those for the non-strange, light quark sea.
New measurements of charged kaon production in deep-inelastic scattering on the deuteron by the HERMES Collaboration \cite{HERMES08} enabled them to present new determinations of the helicity averaged and helicity dependent parton distributions of the strange sea.
Previous to this, the severest constraint on the strange sea distributions has come from di-muon production in neutrino and antineutrino deep inelastic scattering (DIS), mostly from the CCFR \cite{CCFR} and NuTeV \cite{NuTeV} Collaborations. 

The standard method to extract the parton distribution functions (PDFs) of the nucleon is to assume a certain form for the various distributions at an initial scale $Q^2_0 \sim 1 -2$ GeV$^2$ and to perform a global fit  to available experimental data for high energy scattering processes with the help of QCD evolution, which is known to next-to-next-to-leading order (NNLO) in the running QCD coupling constant.
Many groups have been using this method to obtain the PDFs of the nucleon, see for example
\cite{AlekhinKP09,CTEQ6.5S0,CTEQ6.6,MSTW2008,NNPDF1.2,GRV98,GJR08}.
An alternate approach to performing the global fit, mainly developed by the Dortmund group \cite{GRV98,GJR08}, is to assume
only valance-like distributions for the valance quarks, gluons and sea quarks at a somewhat smaller starting scale of $Q_0^2=0.2\sim 0.5$ GeV$^2$.
The PDFs at high $Q^2$ are then mainly dynamically (radiatively) generated from these inputs via evolution.
In all these global fits an initial difference between the strange quark sea and the light quark sea (i.e. SU(3) breaking among the
PDFs of the nucleon) is assumed.
In the standard approach a strangeness suppression factor 
\bea 
r(x) = \frac{s(x) + \sbar{x}}{\dbar(x) + \ubar(x)}
\eea
is introduced, and in the Dortmund group's approach the strange sea is assumed to vanish at the starting scale.
While this initial difference between the strange and light sea improves the $\chi^2$ of the fits, we note that the uncertainties in the final strange and antistrange distributions are much larger than those of the light quark sea.
There is also very little in the theoretical literature to explain this initial difference. 
While it is generally believed that the mass difference between the strange and light quarks is responsible for this difference, the mechanism for generating the difference has not been clearly elucidated.

In the last few years, there has been an extensive effort to fully analyze the available data in order to learn more about the 
strange sea.
The NuTeV data has been analyzed at next-to-leading order (NLO) \cite{MasonNuTeVNLO} and the resulting strange and anti-strange distributions reported.
A global fit to data from the neutrino(antineutrino)-nucleon DIS, 
inclusive charged lepon-nucleon DIS and Drell-Yan experiments, with an emphasis on extracting the strange sea distributions,
was performed by Alekhin, Kulagin and Petti \cite{AlekhinKP09}.
The CTEQ group has studied the magnitude and shape of the strange and antistrange distributions in its global analysis 
\cite{CTEQ6.5S0}, in which the parton distribution function (PDF) set that provided the best fit to the global hard scattering data (CTEQ6.5S0) was presented, along with alternative sets consistent with the data. 
The differences for the strange distribution between CTEQ's best PDF set and their alternative sets are at the $30\%$ level over the range from $x \sim 0.01$ to $x \sim 0.1$.
CTEQ has now published updated PDF sets, CTEQ6.6 \cite{CTEQ6.6}, which are an improvement over the CTEQ6.5 sets 
and more suitable for use in the small $x$ region, although the physics inputs and assumptions about the strange sea distributions are largely preserved.
Updated PDF sets have also been published by the MSTW  \cite{MSTW2008} and NNPDF  \cite{NNPDF1.2} collaborations.


There are two mechanisms responsible for the quark sea production at experimental scales: 
(I) gluons splitting into quark-antiquark pairs,
and (II) non-perturbative contributions such as those from the meson-baryon Fock components of the nucleon wavefunction. 
If the masses of $u$, $d$ and $s$ quarks are light i.e. $m_{q}$ is much smaller than any relevant experimental scale, 
the sea distributions generated through mechanism (I) can be assumed to be flavour independent (SU(3) flavour symmetric) 
$\dbar(x) = \ubar(x) = \sbar(x)$
and quark-antiquark symmetric $\qbar(x) = q(x)$.
On the other hand, sea distributions generated via mechanism (II) can violate these two symmetries. 
Mechanism (II) provides a natural explanation for the observed SU(2) flavour asymmetry among the sea distributions, 
i.e. $\dbar(x) \neq \ubar(x)$ \cite{MST98,MCM98},
and predicts a strange-antistrange asymmetry \cite{SignalT87,BrodskyM96,CaoS99,MM99}.

We note that, while the strange quark is generally believed to be significantly heavier than either $u$ or $d$ 
($m_{s} \sim 0.1$ GeV \cite{Narison06}), this is light compared to experimental scales, and is generally ignored in NLO 
analysis of parton distributions. 
It is also possible that the larger mass of the strange quark could break SU(3) flavour symmetry by restricting the phase space available for $s \sbar$ pairs. 
This phase space volume is proportional to $(Q^2 - 4m_{s}^{2})$, which will restrict the phase space when the virtuality $Q^2$ of the gluon is close to $m_{s}^{2}$ \cite{PeSc}.
However, this scale is much smaller than both the typical experimental scales and the initial scales used in either approach to fitting PDFs. 
Even in a model of the nucleon, such as the MIT bag model or chiral quark soliton model, where PDFs can be calculated at the model scale and evolved up to experimental scales \cite{Adelaide}, the model scales are typically 
$0.2 \sim 0.5\; \mbox{GeV}^2$, which is still significantly larger than $m_{s}^{2}$.
Hence we conclude that mechanism (I), quark pair production from gluons, will not give a large contribution to the breaking of SU(3) flavour symmetry over the range of scales for which a perturbative approach is reasonable.

Thus assuming SU(3) flavour symmetry and quark-antiquark symmetry for the sea distributions generated via mechanism (I), 
we can construct the distribution 
\bea
\Delta (x)=\dbar(x)+\ubar(x)-s(x)-\sbar(x),
\label{Delta}
\eea
which has a leading contribution from mechanism (II), and can be calculated using non-perturbative models describing that mechanism.
$\Delta(x)$ has been previously calculated in the meson cloud model (MCM)  \cite{Sullivan,Thomas83} using covariant perturbation theory and considering only baryon-meson components involving pseudoscalar mesons \cite{SKumano91,FCarvalho00}. 
We present here an updated calculation of $\Delta(x)$ using time-ordered perturbation theory in the infinite momentum frame, which avoids ambiguities around the use of off-mass-shell structure functions \cite{HHoltmannSS}, and considering Fock states involving mesons in the pseudoscalar and vector octets plus baryons in the octet and decuplet.
It is not clear in the MCM where the Fock expansion of the nucleon wavefunction should be truncated, as a rapidity gap to identify the final state is desirable, however, high mass Fock states are naturally suppressed and only make small corrections to the calculation of unpolarized parton distributions.

Combining our calculation for $\Delta(x)$ with results for the non-strange light antiquark sea distributions from global PDF fits
we can calculate the total strange distribution 
$S^+(x)=s(x)+\sbar(x)$
and the strange sea suppression factor  
$r(x)$.

This paper is organized as follows: 
the theoretical formalism used to evaluate $\Delta(x)$, the meson cloud model, is presented in Section 2,  
the numerical results for $\Delta(x)$, the total strange distribution $ S^+(x)$, and the strange sea suppression factor $r(x)$
are presented in Section 3, and Section 4 is our summary and discussions
 
\section{Formalism}

Virtual meson-baryon components are created and annihilated continuously in the nucleon due to energy uncertainty of the system. 
They are relatively long lived, which enables them to contribute to hard processes as these typically have short interaction scales \cite{Sullivan}.
The presence of the meson-baryon components plays an important role in explaining many interesting experimental results, 
including these from polarized and unpolarized DIS and Drell-Yan experiments \cite{MCM98}.
The wave function for the physical nucleon can be written as
\bea
|N\rangle_{\rm physical} &=&  \sqrt{Z} \, |N\rangle_{\rm bare} 
+ \sum_{BM} \sum_{\lambda \lambda^\prime} 
\int dy \, d^2 {\bf k}_\perp \, \phi^{\lambda \lambda^\prime}_{BM}(y,k_\perp^2)  \nonumber \\
& & 
~~~~~~~~~~~~~~~~~  |B^\lambda(y, {\bf k}_\perp); M^{\lambda^\prime}(1-y,-{\bf k}_\perp)
\rangle 
\label{NMCM}
\eea
where the first term is for a `bare' nucleon, $Z$ is the wave function renormalization constant,
and $\phi^{\lambda \lambda^\prime}_{BM}(y,k_\perp^2)$ 
is the wave function of the Fock state containing a baryon ($B$)
with longitudinal momentum fraction $y$, transverse momentum ${\bf k}_\perp$,
and helicity $\lambda$, and a meson ($M$) with momentum fraction $1-y$,
transverse momentum $-{\bf k}_\perp$, and helicity $\lambda^\prime$.
The probability of finding a baryon with momentum fraction $y$ (also known as fluctuation function in the literature)
can be calculated from the wave function $\phi^{\lambda \lambda^\prime}_{BM}(y,k_\perp^2)$,
\bea
f_{BM/N} (y) & = & \sum_{\lambda \lambda^\prime}
\int^\infty_0 d k_\perp^2
\phi^{\lambda \lambda^\prime}_{BM}(y, k_\perp^2)
\phi^{*\,\lambda \lambda^\prime}_{BM}(y, k_\perp^2).
\label{fBMN}
\eea
The probability of finding a meson with momentum fraction $y$ is given by
\bea
f_{MB/N}(y) & = & f_{BM/N} (1-y).
\label{fMBN}
\eea
The wave functions and thereby the fluctuation functions can be derived from effective meson-nucleon Lagrangians
employing time-ordered perturbation theory in the infinite momentum frame \cite{HHoltmannSS}.

The mesons and baryons contribute to a hard scattering process such as DIS via the Sullivan process \cite{Sullivan}, 
where the virtual photon scatters off a meson or a baryon. 
The contribution from the Sullivan processes to the parton distribution functions can be calculated via the 
convolution formula \cite{Thomas83},
\bea
x \delta q(x) & = &  \int^1_x dy f_{BM/N} (y) \left(\frac{x}{y}\right) q_{B}\left(\frac{x}{y}\right) \nonumber \\
& \equiv & f_{BM/N} \otimes q_B  \label{xdeltaq} \\
x \delta \qbar(x) 
& \equiv & f_{MB/N} \otimes \qbar_M
\label{xdeltaqbar}
\eea
where $q_B$ and $\qbar_M$ are valence parton distributions in the baryon and meson.
Eqs.~(\ref{xdeltaq}) and (\ref{xdeltaqbar}) represent the contribution from the process where 
the baryon and meson respectively participate in the hard process 
(while the other partner in the baryon-meson component is a spectator).

In a recent paper, Strikman and Weiss \cite{SW09} investigated the dependence on impact parameter of such baryon-meson 
cloud contributions to sea distributions.
They found that about one third of the SU(2) symmetry breaking distribution $\dbar(x) - \ubar(x)$ arises from large distance 
contributions, and a similar fraction of $\Delta(x)$ is due to scattering at large impact parameter, where they define large distance as greater than the nucleon's transverse axial charge radius, estimated at 0.55 fm.
We would argue that this approach is contrary to models of nucleon structure such as the Cloudy Bag Model (CBM) \cite{CBM} and the chiral quark soliton model {$\chi$QSM} \cite{XQSM}, where quarks couple to pions and other mesons throughout the volume of the nucleon, not just near the surface.
Also the Sullivan mechanism distinguishes states by their rapidity not impact parameter, and it is not surprising that significant contributions to sea distributions can arise from {\em e.g.} fast moving mesons close to the centre of the nucleon as well as those far from the centre.
The rapidity gap necessary for identifying MCM contributions can be seen in the cross-section for leading neutron production in DIS on protons at HERA \cite{MCMneutrons}.

The leading contributions to the distribution $x \Delta (x)$ come from non-perturbative processes
(see Eqs.~(\ref{xdeltaq}) and (\ref{xdeltaqbar})). 
The Fock states we consider include
$\left | N \pi \right >, \left | N \rho \right >, \left | \omega N \right >, \left | \Delta \pi \right >, \left | \Delta \rho \right >$, 
$\left | \Lambda K \right >, \left | \Lambda K^* \right >,  \left | \Sigma K \right >$, and $\left | \Sigma K^* \right >$.
We obtain the following expression for $x \Delta (x)$
\bea
x\Delta (x) &=& 
\left( f_{\pi N/N}+f_{\pi\Delta/N}+f_{\rho N/N}+f_{\rho\Delta/N}+f_{\omega N/N} \right) \otimes V_\pi  
\nonumber \\
	& &   - \left[   \left( f_{\Lambda K/N}  + f_{\Lambda K^*/N} \right) \otimes s^\Lambda
			 +\left( f_{\Sigma K/N}  + f_{\Sigma K^*/N} \right) \otimes s^\Sigma \right. \nonumber \\
	& &	 
	\left.  +\left( f_{K \Lambda/N}+f_{K \Sigma /N} +f_{K^* \Lambda/N}+f_{K^* \Sigma /N} \right) \otimes \sbar^K \right] .
\label{Delta_MCM}
\eea
In the above expression we have taken the unpolarized parton distributions for the $\rho$, and $\omega$ to be the same as that for 
the $\pi$ ($V_{\pi}(x)$), for which we use the Gl\"{u}ck-Reya-Schienbein paramaterization \cite{GRS99}. 
Also we have taken the unpolarized parton distributions for the $K$ and $K^*$ to be equal ($ \sbar^{K}(x)$). 
This distribution, and the baryon valence distributions $ s^{\Lambda}(x) $, $s^{\Sigma}(x) $ are calculated
using a variant of the MIT bag model developed by the Adelaide group \cite{Adelaide} and
ourselves \cite{CaoS01,CaoS03,BisseyCS06,Signal97}.
These distributions are calculated at the model scale of $\mu_0^2=0.23$~GeV$^2$ and 
evolved to $Q^2=2.5$ GeV$^2$ using the programme provided in \cite{HiraiKM98,MiyamaK96}.

The expressions for the fluctuation functions involved are presented in Appendix A. The calculations details for the
fluctuation functions can be found in \cite{CaoS99,HHoltmannSS,CaoS01,CaoS03}.
We note that all the relevant coupling constants are determined by experiment and SU(3) symmetry, and the form factors 
used to make relevant momentum integrals finite are well constrained by fits to DIS data.  

The non-strange light quark sea distributions are well determined by the global PDF fits to all available experimental data.  
Combining the global fit results for $\dbar(x)+\ubar(x)$ with our calculation for  $\Delta(x)$ we are able to estimate the strange sea distributions via
\bea
x\left[ s(x)+\sbar(x)\right]= x\left[ \dbar(x)+\ubar(x)\right]_{\mbox{Fit}} - x\Delta(x),
\label{xS+}
\eea
and the strange sea suppression factor
\bea
r(x)=1- \frac{\Delta(x)}{\left[\dbar(x)+\ubar(x)\right]_{\mbox{Fit}}}.
\eea

\section{Results}

The numerical results  for $x\Delta(x)$ are shown in Fig.~1.  
The largest contributions to the light antiquark sea ($\dbar$ and $\ubar$) come from the Fock states $\left | N \pi \right >, \left | N \rho \right >$ and 
$ \left | \Delta \pi \right >$, while the Fock states involving $K$ and $K^*$ mesons and responsible for the strange sea
are of roughly equal magnitude.
We also show the results when Fock states containing $K^{*}$ mesons are omitted in order to provide some estimate of the uncertainties in the model calculation. 
We can see that this has a small effect on the calculation of $x\Delta(x)$, and gives us some confidence that the uncertainties in the MCM calculations are under control.
For comparison we show the HERMES measurements for $x(s+\sbar)$
(which are obtained using leading-order analysis)  
combined with the CTEQ group's leading order PDFs (CTEQ6L) for $x(\dbar+\ubar)$ to obtain 
$x\Delta(x)$ as data points. 
We also show 
a similar combination of the NuTeV NLO parametrisation of $x(s+\sbar)$ \cite{MasonNuTeVNLO} 
with the CTEQ6M PDF set for $x(\dbar+\ubar)$, and 
results from the CTEQ6.5 \cite{CTEQ6.5S0}, CTEQ6.6M \cite{CTEQ6.6}, and MSTW2008 \cite{MSTW2008} PDF sets.
All results are given at $Q^2=2.5$ GeV$^2$.
The shaded area represents the allowed range for $x\Delta(x)$ consistent with the uncertainty in the $xS^+(x)$ distribution given by the CTEQ6.5 PDF set \cite{CTEQ6.5S0}.
This was determined by applying $90\%$ confidence criteria to the di-muon production data sets which required that the momentum fraction carried by the strange sea to be in the range of $0.018 \rightarrow 0.040$.

It can be seen that our calculated $x\Delta(x)$ is significantly smaller than the distribution calculated using MSTW2008 PDF set,
the `experimental' distribution based on NuTeV data, and the central values of the CTEQ6.5 distributions in the region $x < 0.2$, 
while there is reasonably good agreement with the HERMES results except for the  region around $x \sim 0.10$.
Our calculation agrees well with the $x\Delta(x)$ obtained using the CTEQ6.6 PDF set in the region $ x < 0.2$.
We note that the two `experimental' distributions based on HERMES and NuTeV data combined with CTEQ parameterisations of 
$x(\dbar+\ubar)$ do not agree well. 
It would be very interesting to see if this difference remains when the HERMES data is analyzed at NLO in QCD. 
We note that the MCM calculation of $x\Delta(x)$ is independent of any global PDF set for the proton.

\begin{figure}		
\begin{center}
 \includegraphics[width=100mm]{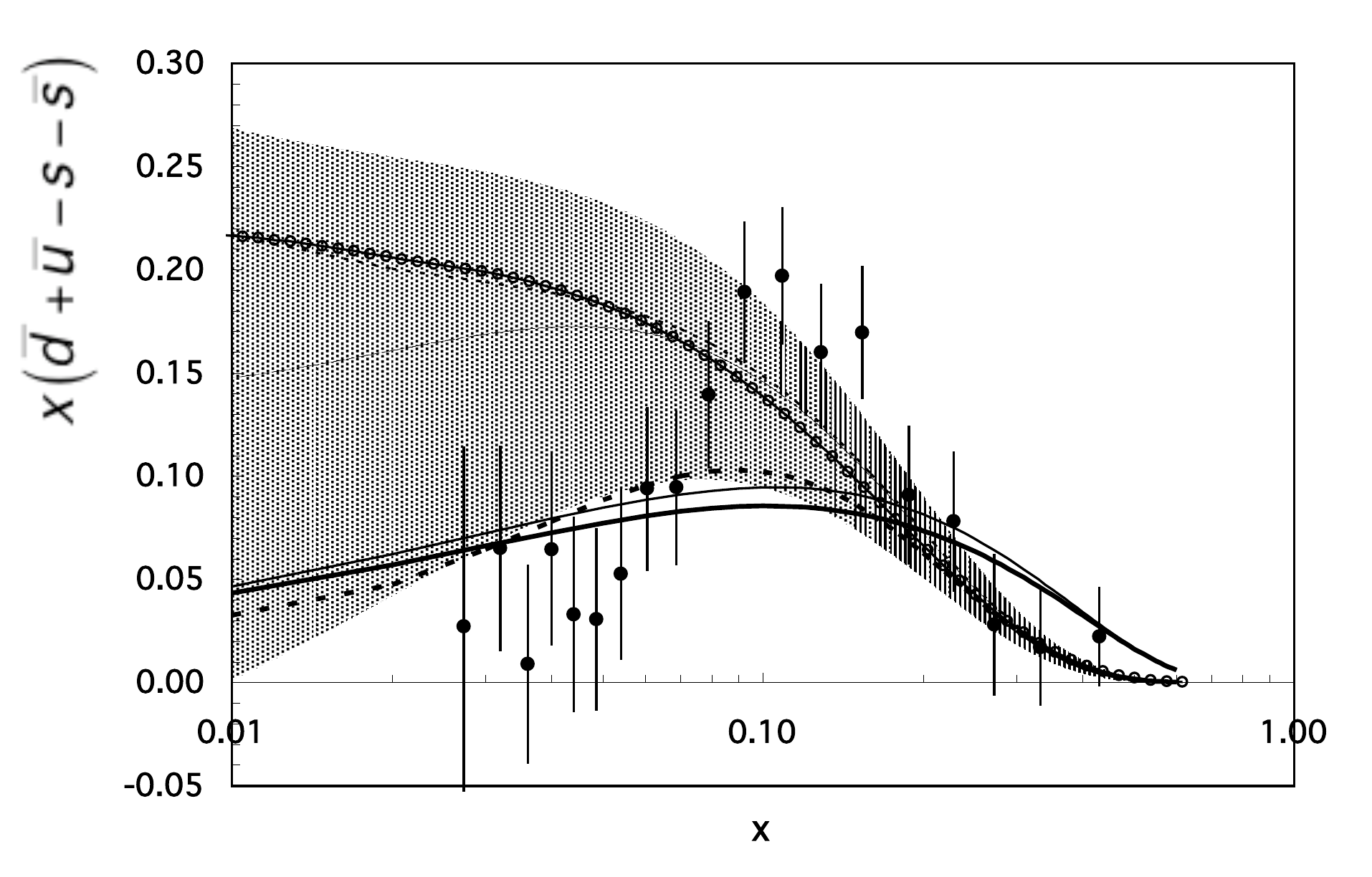}
\vspace{0.0cm}
\caption{A comparison of the MCM calculations for $x\Delta(x)$ (the thick solid curve - with $K^*$ contribution; the thin solid curve - without $K^*$ contribution)  with the `experimental' distributions based on HERMES \cite{HERMES08} (the data points)  and NuTeV \cite{MasonNuTeVNLO} 
(the solid curve with open-circle markers) data.
The results from CTEQ6.6M  \cite{CTEQ6.6} (the thick dashed curve), MSTW2008 \cite{MSTW2008} (the dashed curve) and
CTEQ6.5 \cite{CTEQ6.5S0} (the shaded area with the middle curve giving the central values) are also shown.}
\end{center} 
\end{figure}

The results for the total strange and antistrange distributions are shown in Fig.~2.
For these calculations we combined our calculation of $x\Delta(x)$ with the $(\dbar +\ubar)(x)$ distribution from the CTEQ6.6M set. 
We find that our calculations are in good agreement with the CTEQ6.6M set, and with the HERMES data for the region $x < 0.07$,
but are larger than those from the MSTW2008 and CTEQ6.5 sets, and the NLO analysis of NuTeV data \cite{MasonNuTeVNLO}.
Our calculation for $x(s+\sbar)$ becomes negative for $x > 0.25$ which is unreasonable. 

The reason for this could be that our model calculations overestimate $x\Delta(x)$, or that
$x\left(\dbar(x)+\ubar(x) \right )$ is underestimated in the CTEQ6.6 set, or both.
Our calculations are sensitive to the calculated PDFs we have used for the baryons ($\Delta$, $\Lambda$ and $\Sigma$),
and at large $x$ these calculations become unreliable owing to the kinematic limitations of the MIT bag model.
After convolution with the fluctuation functions, this uncertainty can affect the calculated $x\Delta(x)$ and strange
distributions in the medium and large $x$ regions, especially as these distributions are calculated in terms of differences between
the model PDFs. We also note that omitting the $K^{*}$ states from the calculation makes almost no difference in this region.

\begin{figure}		
\begin{center}
\includegraphics[width=100mm]{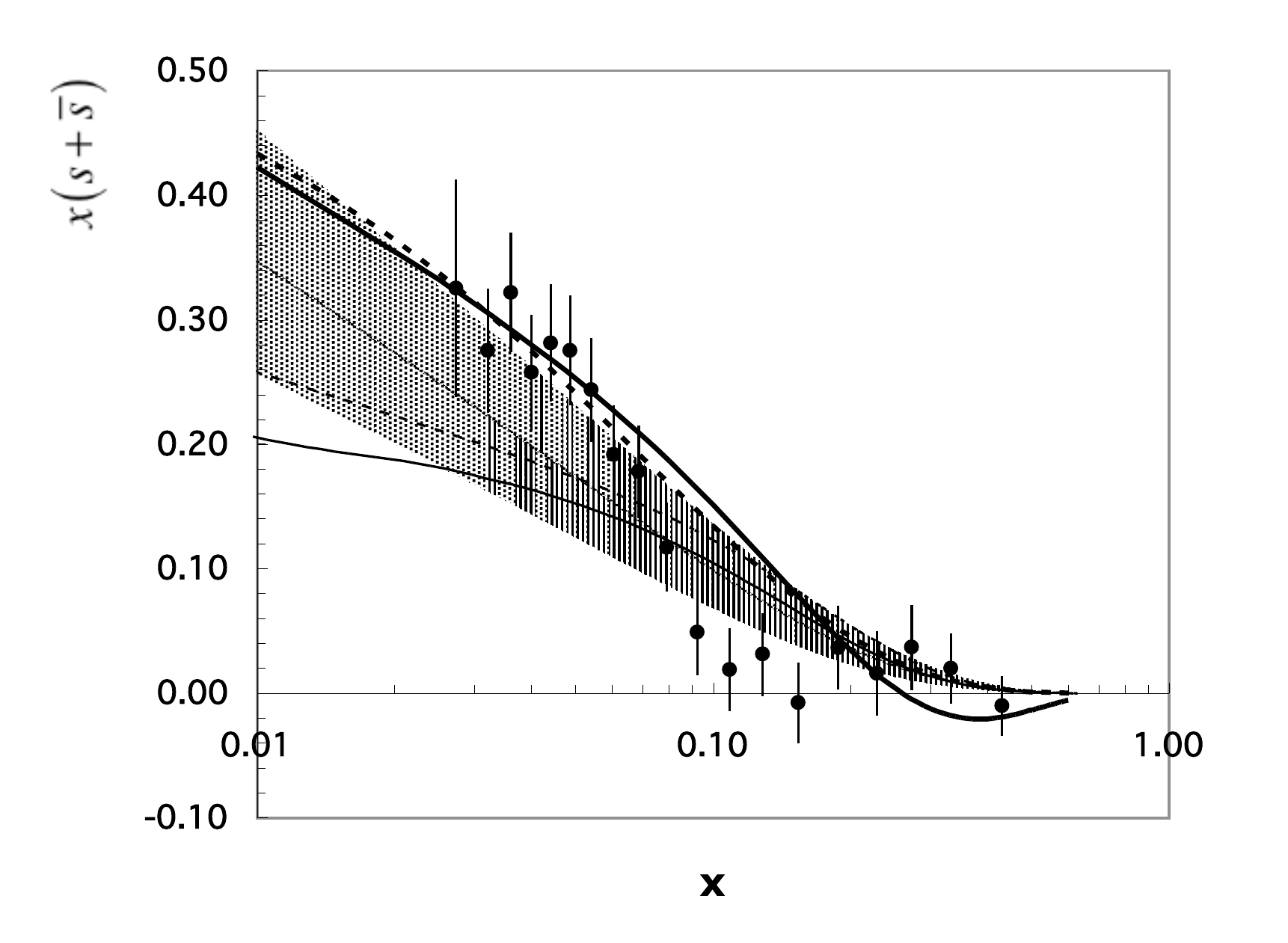}
\vspace{0.0cm}
\caption{The sum of the strange and antistrange quark distributions, shown as $x(s+\sbar)$, from the MCM calculations (the thick solid curve),
the HERMES measurements \cite{HERMES08} (the data points), and the next-to-leading order analysis of NuTeV dimuon data \cite{MasonNuTeVNLO}
(the thin solid curve).
The global fit results from CTEQ6.6M \cite{CTEQ6.6} (the thick dashed curve), 
MSTW2008  \cite{MSTW2008} (the dashed curve) and CTEQ6.5 \cite{CTEQ6.5S0} 
(the shaded area with the middle curve giving the central values) are also shown.}
\end{center} 
\end{figure}

Our calculation of the strangeness suppression $r(x)$ is shown in Fig.~3. 
Again, we have combined the results from HERMES and NuTeV with CTEQ $(\dbar+\ubar)$ sets at the appropriate order to find 
`experimental' determinations of this ratio.
The MCM calculation gives a result that agrees reasonably well with HERMES and the CTEQ6.6M set for $x<0.07$, and 
is larger than those obtained in the previous global PDF fits,
which suggests that the suppression of strange sea relative to the 
non-strange light antiquark sea is not as large as has been previously believed.

\begin{figure}	
\begin{center}
 \includegraphics[width=100mm]{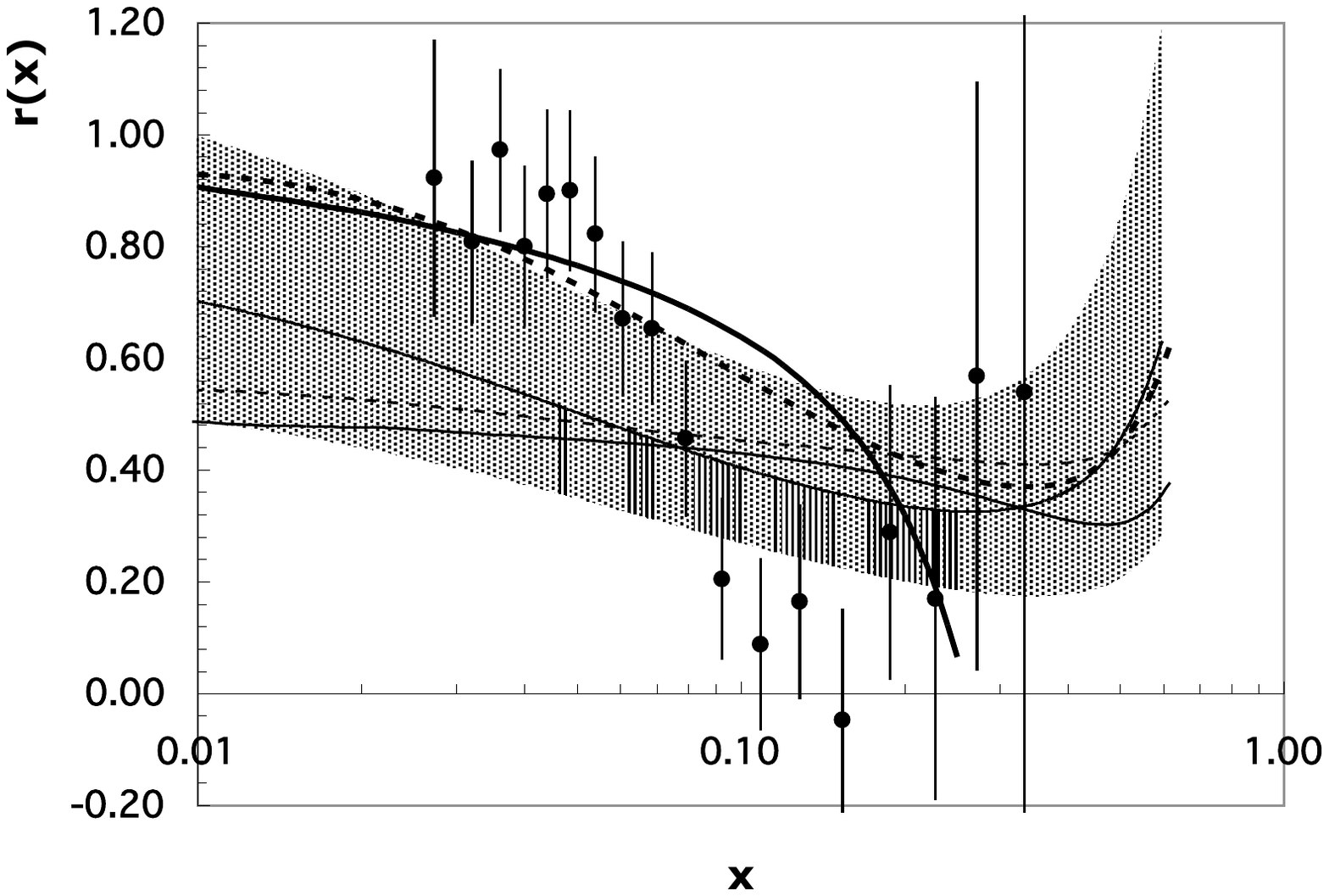}
\vspace{-0.0cm}
\caption{The relative strength of strange and antistrange distributions to the non-strange light antiquark distributions, $r(x)$:
obtained using the MCM calculations for $x(s+\sbar)$ and CTEQ6.6M \cite{CTEQ6.6} results for $x(\dbar+\ubar)$ (the thick solid curve),
the HERMES \cite{HERMES08} measurements for $x(s+\sbar)$ and CTEQ6L \cite{CTEQ6.6} set for $x(\dbar+\ubar)$ (the data points), and 
the next-to-leading order analysis of NuTeV dimuon data giving $x(s+\sbar)$
\cite{MasonNuTeVNLO} and CTEQ6M \cite{CTEQ6.6} set for $x(\dbar+\ubar)$ (the solid curve).
The global fitting results from CTEQ6.6M \cite{CTEQ6.6} (the thick dashed curve), MSTW2008 \cite{MSTW2008} (the dashed curve) and 
CTEQ6.5 \cite{CTEQ6.5S0}  (the shaded area with the middle curve giving the central values) are also shown.}
\end{center} 
\end{figure}

\section{Summary and discussions}

For light quarks, non-perturbative processes are the dominant mechanism for the difference $\Delta(x)$ between the 
non-strange antiquark distributions and the strange sea distributions, 
since the perturbative process of gluons splitting into quark and antiquark pairs is both flavour symmetric and quark-antiquark symmetric at leading order.
We have calculated this difference using the meson cloud model and estimated the
total strange plus antistrange distribution and the strange distribution suppression factor 
by combining our calculations for $\Delta(x)$
with the non-strange light antiquark distributions determined from global parton distribution functions fits.
Our calculations for the strange sea distributions agree with the HERMES measurements and that calculated using 
the CTEQ6.6 PDF set, but are larger than the NuTeV data and those obtained using the CTEQ6.5 and MSTW2008 sets. 
There appears to be a large discrepancy between the HERMES and NuTeV measurements in the region $x <  0.05$, however, distributions in this region are sensitive to the details of the experimental analysis. 
In particular, analysis of the HERMES data at NLO would be required in order to check the compatibility of the two 
measurements of the strange sea distributions.

This calculation could be extended to helicity dependent sea distributions.
We note that the helicity dependent strange distributions $x\Delta s(x) + x\Delta \sbar(x)$ has been measured by the HERMES
group \cite{HERMES08,HERMES05},
and found to be somewhat smaller in magnitude than expected on the basis of $SU(3)_{f}$ arguments \cite{Leader06}.
Previous MCM calculations \cite{BisseyCS06} have been in accord with the HERMES data \cite{HERMES05},
however a calculation analogous to the calculation presented here could give added confidence in these results.
One drawback of this method would be that the current data on the helicity dependent non-strange sea distributions has much larger uncertainties than
the helicity independent data which (along with those of the model calculation) may well be larger than any small helicity dependent strange sea distribution.

\appendix
\section{Fluctuation functions}
The fluctuation function for $N \ra B M$ can be written as
\bea
f(y,k_T) = \frac{m_N m_B}{4 \pi^2 y (1-y)} \frac{G^2(y,k_T)}{\left(m_N^2-m_{BM}^2\right)^2} V^2(y,k_T),
\eea
where $m_N$, $m_B$, and $m_M$ are the mass for the nucleon, baryon and meson, respectively,
and $m_{BM}^2$ is the invariant mass squared of the baryon-meson Fock state,
\bea
m_{BM}^2(y,k_T) &=& \frac{\left( m_B^2+k_T^2 \right)}{y}+\frac{\left(m_M^2+k_T^2 \right)}{1-y}.
\eea
The vertex includes a form factor $G$
\bea
G(y,k_T) &=& {\rm Exp}\left[ (m_N^2-m_{BM}^2)/(2 \Lambda_{cut}^2)\right],
\eea
with $\Lambda_{cut}$ being a cut-off parameter.
In this work we take $\Lambda_{\mbox{oct}} = 0.8$ GeV and $\Lambda_{\mbox{dec}} = 1.0$ GeV for fluctuations  involving 
octet and decuplet baryons respectively. 
These values of the cut-offs give a good description of a wide range of DIS distributions involving MCM contributions 
\cite{MST98,CaoS99,CaoS01,CaoS03,BisseyCS06}.

For the fluctuations $N \ra N\pi$, $\Lambda K$, and $\Sigma K$,
\bea
V^2(y,k_T) = 3 g^2\frac{ k_T^2+({m_B}-{m_N} y)^2}{4 {m_B} {m_N} y},
\eea
where the coupling constants are taken to be $g_{N\pi/N}=13.07$, $g_{\Lambda K/N}=-13.98$,
 and $g_{\Sigma K/N}=2.69$ \cite{HHoltmannSS,MachleidtHE87}.
For the fluctuations $N \ra \Delta \pi$,
\bea
V^2(y,k_T) = 2 f^2 \frac{ \left[{k_T}^2+({m_B}-{m_N} y)^2\right] \left[{k_T}^2+({m_B}+{m_N} y)^2\right]^2}{24{m_B}^3 {m_N} y^3},
\eea
where the coupling constant $f_{\Delta \pi/N}=12.43$ GeV$^{-1}$ \cite{HHoltmannSS,MachleidtHE87}.
For the fluctuations $N \ra N \rho$, $N \omega$, $\Lambda K^*$, and $\Sigma K^*$,
\bea
V^2(y,k_T) &=& 3  \frac{1}{{4 {m_B} {m_N} y^3}} \left\{
\frac{8 f^2 y^2 k_T^4}{(y-1)^2}
+\frac{2 \left[ g+2 f {m_N} (y-1) \right]^2 y^2 k_T^2}{(y-1)^2} \right. \nonumber \\
&&\hspace{0.5cm} +\frac{2 \left[ g y-2 f {m_B} (y-1) \right]^2 y^2 k_T^2}{(y-1)^2}  \nonumber \\
&&\hspace{0.5cm}+k_T^2 \left[ \frac{g ({m_B}-{m_N}) (y-1) y}{{m_M} (y-1)} \right. \nonumber \\
&&\hspace{1.5cm}\left. - \frac{f (y+1) \left[k_T^2-{m_B}^2 (y-1)+y \left({m_N}^2 (y-1)-{m_M}^2\right)\right]}
{{m_M} (y-1)} \right]^2 \nonumber \\
&& \hspace{0.5cm}+ y^2 \left[ \frac{ g \left[k_T^2+{m_B} {m_N} (y-1)^2-{m_M}^2 y\right]}
{{m_M} (y-1)} \right.  \nonumber \\
&&\hspace{1.5cm}\left. +\frac{f ({m_N} y-{m_B}) \left[k_T^2+{m_B}^2+{m_N}^2 y^2-\left({m_B}^2+{m_M}^2+{m_N}^2 \right) \right]}
{{m_M} y} \right]^2 \nonumber \\
&& \hspace{0.5cm} +2 y^2 \left[ \frac{ g(y-1) ({m_B}-{m_N} y)}{(y-1)} \right. \nonumber \\
&& \hspace{1.5cm} \left. \left. +\frac {2 f \left[k_T^2-({m_B}+{m_N}) (y-1) ({m_B}-{m_N} y)\right]}{(y-1)} \right]^2
\right\}.
\eea
where the coupling constants are taken to be $g_{N \rho /N}=3.25$, $f_{N \rho /N}/g_{N \rho /N}=6.1/4m_N$,
$g_{N \omega/N}=10.09 $, $f_{N \omega /N}/g_{N \omega /N}=0$,
$g_{\Lambda K^*/N}=-5.63$, $f_{\Lambda K^*/N}=-4.89$ GeV$^{-1}$, 
$g_{\Sigma K^*/N}=-3.25$, and $f_{\Sigma K^*/N}=2.09$ GeV$^{-1}$  \cite{HHoltmannSS,MachleidtHE87}.
For the fluctuations $N \ra \Delta \rho$,
\bea
V^2(y,k_T) &=& \frac{f^2}{6{m_B}^3 {m_N} (y-1)^2 y^3}
\left\{k^6+\left[\left(4 y^2-4 y+3\right) {m_B}^2+\left(2{m_M}^2+{m_N}^2\right) y^2\right] k^4  \right. \nonumber \\
&& 
+\left[(y-1)^2 \left(3 y^2-2 y+3\right){m_B}^4 +2 y^2 \left[\left(y^2+2\right) {m_M}^2
+2 {m_N}^2 (y-1)^2\right]{m_B}^2 \right. \nonumber \\
&& \left.
+{m_M}^2 \left({m_M}^2+2 {m_N}^2\right) y^4\right] k^2 \nonumber \\
&& 
+{m_M}^4{m_N}^2 y^6+{m_B}^6 (y-1)^4+{m_B}^2 {m_M}^2 \left[3 {m_M}^2+2 {m_N}^2
(y-1)^2\right] y^4 \nonumber \\
&& \left.
-12 {m_B}^3 {m_M}^2 {m_N} (y-1)^2 y^3+{m_B}^4 \left[2
{m_M}^2+3 {m_N}^2 (y-1)^2\right] (y-1)^2 y^2\right\},
 \eea
where the coupling constant $f_{\Delta \rho/N}=20.82$ GeV$^{-1}$  \cite{HHoltmannSS,MachleidtHE87}.



\end{document}